\def\Murcia{Departamento de Matem\'atica Aplicada, Facultad de Inform\'atica, Campus de 
    Espinardo, 30100 Murcia, Spain} 
 
\def\CarlosI{Instituto de F\'\i sica Te\'orica y Computacional Carlos I, 
Facultad de Ciencias, Universidad de Granada, Campus de Fuentenueva, Granada 18002, 
Spain} 
 
\def\IAA{Instituto de Astrof\'{\i}sica de Andaluc\'{\i}a, Apartado Postal 3004, Granada 
       18080, Spain} 
\def\Comision{Work partially supported by the DGICYT.}

\def \ba {\begin{array}}
\def \ea {\end{array}}

\def \bea {\begin{eqnarray}}
\def \eea {\end{eqnarray}}

\def \be {\begin{equation}}
\def \ee {\end{equation}}

\def\ni{\noindent}
\def\nn{\nonumber}

\def\l[{\left[}
\def\r]{\right]}

\def\t{\tilde{X}_t}

\documentstyle[11pt]{article}

\begin{document}

\begin{center} 
{\LARGE {\bf New insights in particle dynamics from group cohomology}} 
\end{center} 
 
\bigskip 
\bigskip 
 
\centerline{V. Aldaya$^{2,3}$, J.L. Jaramillo$^{2,3}$ 
       and J. Guerrero$^{2,3,4}$ } 
 
\bigskip 
 
\footnotetext[1]{\Comision} 
\footnotetext[2]{\IAA} \footnotetext[3]{\CarlosI} 
\footnotetext[4]{\Murcia} 
 
\bigskip 
 
 
\small 
\setlength{\baselineskip}{12pt} 
 
\begin{list}{}{\setlength{\leftmargin}{3pc}\setlength{\rightmargin}{3pc}} 
\item The dynamics of a particle moving in background 
electromagnetic and gravitational fields is revisited from a Lie 
group cohomological perspective. Physical constants characterising the 
particle appear as central extension parameters of a group which is obtained 
from a centrally extended kinematical group (Poincar\'e or 
Galilei) by 
making {\it local} some subgroup. The corresponding dynamics is generated by a
vector field inside the kernel of a presymplectic form which is derived from
the canonical left-invariant one-form on the extended group. A 
{\it non-relativistic} limit is derived from the {\it geodesic} motion 
via an In\"on\"u-Wigner contraction. A deeper analysis of the 
cohomological structure reveals the possibility of a new force associated with
a non-trivial mixing of gravity and electromagnetism leading to in principle
testable predictions.
\end{list} 
 
\normalsize 

\vskip 1cm

\section{General setting}
The aim of this work is that of clarifying 
the underlying algebraic structure behind the dynamics of a particle moving 
inside a background field. We shall see how the constants characterizing the 
properties of the particle and its couplings can be understood in terms of
the parameters associated with the central extensions of certain groups,
thus bringing into scene group-cohomological concepts.

In order to motivate the role of central extensions, let us firstly recall 
some basic and well-known facts with the help of an example representing, 
perhaps,
the simplest physical system one can imagine: the free particle with 
Galilei symmetry. In order to define the system we make use of the
{\it Poincar\'e-Cartan} form defined on what we shall call {\it evolution} 
space contructed from phase space by adding time, $(x,p,t)$,
\bea
\Theta_{PC}=pdx-\frac{p^2}{2m}dt \ \ . 
\eea
\pagebreak
A realization of Galilei group, parametrized by $(b,a,V)$ (we omit
rotations) and the corresponding infinitesimal transformations (generators), 
on this evolution space is the following:\\
\be
\begin{array}{lll} &x'=x+a+Vt&\;\;\;\;\;\;\;
\;
X_a=
\frac{\partial}{\partial x} \\
&t'= t+b&\;\;\;\;\;\;\;\;
X_b=\frac{\partial}{\partial t} \\
&p'=p+mV&\;\;\;\;\;\;\;\; X_V=m
\frac{\partial}{\partial p}+t\frac{\partial}{\partial x}\end{array} 
\label{galnonext} 
\ee
When checking the invariance of the Poincar\'e-Cartan form under the
Galilei group we realize that its variation under the action of 
{\it boosts} is a total differential, rather than zero. In infinitesimal 
terms, the Lie derivative
\bea 
L_{X_V}\Theta_{PC}=d(mx)\neq 0\;\; , 
\eea
thus leading to the idea of semi-invariance. Of course, this is not a problem 
at the classical level, since the equations 
of motion are not sensitive to such a total-derivative  variation.
Despite it is not necessary in a strict manner, let us raise to the level of a 
postulate the claim for strict invariance
and let us see what consequences we can derive from this assumption.

In order to achieve such strict invariance, let us extend the 
evolution space with a new 
variable $\eta=e^{i\phi}$ which transforms under the Galilei group in such a 
way that the  variation of the total differential of $\phi$ 
inside a modified one-form,
$\Theta\equiv\Theta_{PC}+d\phi$,  cancels out the term $d(mx)$. That is:
\bea
d\phi'=d\phi-d(mx) 
\eea
The corresponding finite action of the Galilei group on this new variable is
then (a new group parameter $\varphi$ must be included for the following
expression to be a proper action):
\bea
\eta'=\eta e^{-i[\frac{1}{2}m V^2t+mVx+\varphi]}\;\; \label{accioncent},
\eea
which together with the action on the rest of the variables in the extended 
evolution space (\ref{galnonext}), allow us to compute the infinitesimal 
generators as well as their commutation relations: 
\bea
[\tilde{X}_b,\tilde{X}_a]=0\;\;,\;\,[\tilde{X}_b,\tilde{X}_V]=
\tilde{X}_a\;\;,\;\;[\tilde{X}_a,\tilde{X}_V]=m\tilde{X}_\varphi \ \ . 
\eea
We notice the presence of a central term in the last commutator and, 
therefore, that
the claim for strict invariance has led us to the centrally extended Galilei
group.

The crucial fact about this phenomenon leading to the central extension
is that it is not related to a particular realization 
of the group, but it is a consequence of its intrinsic algebraic structure, 
in fact related to group-cohomologial features.
This suggests to consider the group itself (a centrally extended group,
in fact) as the starting point in the 
definition of the dynamics of a physical system.

This is exactly the aim of the so-called Group Approach to Quantization (GAQ)
(see \cite{GAQ} and references there in) which tries to derive a dynamics 
directly from the strict symmetry of the corresponding physical system, the  
group-cohomology playing a central role. Even though the main stress of the 
approach leans on its quantum aspects, it also has non-trivial implications
at the classical level, which are the aspects we are going to emphasize 
here. The extended evolution space and the Poincar\'e-Cartan form
are generalized by objects that can be completely recovered from the 
centrally extended
symmetry group. The latter has the structure of a principal fibre bundle 
$\tilde{G}$ whose base is the non-extended group $G$ and where the fibre
is the $U(1)$ group of phase invariance of Quantum Mechanics. The relevant 
cohomology in the construction is that of $G$.
The object generalizing the Poincar\' e-Cartan form is that component
of the left-invariant canonical one-form on the group which is dual to
the vertical (or central) vector field: $\Theta={\theta^L}^{(\zeta)}$.
By its own construction, this $\Theta$ generalizing $\Theta_{PC}+d\phi$ 
(which we will call quantization one-form) is invariant
under the left action of the extended group (meanwhile $\Theta_{PC}$ is
semi-invariant under $\tilde{G}/U(1)$).
Regarding classical dynamics, the form $d\Theta$ can be seen as a 
pre-symplectic form, in such a way that the solution space (i.e. the phase 
space) is obtained from the group by getting rid of those variables inside 
the kernel of $d\Theta$. In this way, the trajectories of the vector fields
inside this kernel can be seen as generalized equations of motion 
\cite{Godbillon}. In 
principle, there is a certain ambiguity in the choice of the Hamiltonian 
vector field, but this problem will not arise in the system we shall 
consider here.

Let us illustrate this technique revisiting the free galilean particle
under this perspective.
For instance, from the realization of the group on the evolution phase 
(\ref{galnonext}) and (\ref{accioncent}), we can derive the following
group law for the centrally extended Galilei group:
\be
\begin{array}{rclrcl}
b"&=&b'+b \ \ \ \ \ \ \ \
&a"&=&a'+a+V'b \\
V"&=&V'+V &
\zeta"&=&\zeta'\zeta e^{-i\frac{m}{\hbar}[V'a+\frac{1}{2}
b{V'}^2)]} \ \ .
\end{array}\label{galgl}
\ee
We can use it to compute the right and left invariant vector fields (from now
on, we shall write $x$ for $a$, $t$ for $b$, $v$ for $V$, $\varphi$ for
$\phi$ and therefore $\zeta$ for $\eta$, whenever the 
discussion takes place in the GAQ setting, and set $\hbar=1$)
\be
\begin{array}{ll}
\tilde{X}^L_t=\frac{\!\!\partial}{\partial t}+v\frac{\!\!\partial}
{\partial x}
-\frac{1}{2}mv^2\frac{\!\!\partial}{\partial\varphi}
\;\;\;&\tilde{X}^R_t=\frac{\!\!\partial}{\partial t}\\
\tilde{X}^L_x=\frac{\!\!\partial}{\partial
x}-mv\frac{\!\!\partial}{\partial \varphi}&\tilde{X}^R_x=
\frac{\!\!\partial}{\partial x} \\
\tilde{X}^L_v=\frac{\!\!\partial}{\partial v}
&\tilde{X}^R_v=
\frac{\!\!\partial}{\partial v}+t\frac{\!\!\partial}{\partial x} 
-mx\frac{\!\!\partial}{\partial\varphi}\\
\tilde{X}^L_\varphi=\frac{\!\!\partial}{\partial\varphi}&
\tilde{X}^R_\varphi=\frac{\!\!\partial}{\partial\varphi}
\end{array}\label{invvect} \ \ ,
\ee
and the quantization one-form
\bea
\Theta\equiv{\theta^L}^\varphi=mvdx-\frac{1}{2}mv^2dt+d\varphi \ \ ,   
\eea
where the first two terms in the r.h.s. correspond to the original 
Poincar\'e-Cartan form.

\section{Interactions}
\subsection{Electromagnetism}
Up to now we have sketched two alternative ways of considering the role of 
the symmetry group when dealing with the dynamics of a physical system. The
first one make use of a particular realization of the group on a extended
phase space, while the second one (that of GAQ) emphasizes the singular 
role of the group, considering it as the departing point in the analysis.
At this point we switch on interactions, starting with the
electromagnetic force, and study the 
situation from the perspective of both technical strategies.

A natural question arising when considering the central 
extended Galilei group refers to the consequences of turning into 
local the $U(1)$ part of the symmetry \footnote{Intuition is led
by the fact that a local $U(1)$ group is intrincally related to 
electromagnetism by means of a minimal coupling principle \cite{Utiyama}.}. 
Dwelling on the 
setting of the first approach \cite{AA83}, the Lie algebra with a local $U(1)$
is composed by the former Galilei 
generators realized on the extended evolution space, together with the tensor 
product of local functions and the central term: $f(x,t)\otimes X_\phi$.
But when we check the invariance of the modified Poincar\'e-Cartan form 
($\Theta_{PC}+d\phi$) under
the new generators, semi-invariance reappears into scene:
\bea
L_{f\otimes X_\phi}\Theta=df 
\eea
We follow here the same strategy as before, i.e. we look for new
variables extending the evolution space whose variation under the symmetry 
group compensates the variation $df$ in a newly 
modified one-form $\Theta$. 
Fortunately in this case there are natural guesses and we are able to find 
new variables $A_0, A_x$, transforming in the desired way ($A'=A-df$).

\ni The realization of the fields in the newly extended phase space 
($(x,p,t,\phi,A_0,A_x)$) is:
\be
\begin{array}{l}
\tilde{X}_b=\frac{\partial}{\partial t} \\
\tilde{X}_a=\frac{\partial}{\partial x} \\
\tilde{X}_v=t\frac{\partial}{\partial x}+m\frac{\partial}{\partial p}-mx
 \frac{\partial}{\partial \phi}+ A_x\frac{\partial}{\partial A_0}\\
\widetilde{f\otimes X_\phi}=-f\frac{\partial}{\partial \phi}-
\frac{\partial f}{\partial x}
\frac{\partial}{\partial A_x}+
\frac{\partial f}{\partial t}\frac{\partial}{\partial A_0}
\end{array}\label{realelec}
\ee
The new strictly invariant one-form is:
\bea
\Theta=pdx-\frac{p^2}{2m}dt-A_xdx+A_0dt +d\phi 
\eea
\ni from which the Lorentz force felt by the particle can be straightforwardly
derived. 

As an alternative approach, we apply the techniques of GAQ to this problem, 
which results in a
more algorithmic and general treatment. In fact, if we
consider an arbitrary group $\tilde{G}$ whose infinitesimal generators
are $\{X_\alpha\}, \;(\alpha=1,...,n)$ and an invariant subgroup 
$\{X_i\},\;(i=1,...,m<n)$,
we can make {\it local} the latter, obtaining an algebra 
spanned by $\{f\otimes X_{i},\;X_\alpha\}$
and whose new commutators are:
\bea
\left[X_\alpha,\;f\otimes X_{i}\right]=
f\otimes\left[X_\alpha,\;X_{i}\right]+
L_{X_\alpha}f\otimes X_{i}
=f\otimes C^{j}_{\alpha i}
X_{j}+L_{X_\alpha}f\otimes X_{i}
\eea
One then applies GAQ tools to obtain the quantization
one-form $\Theta$.

In the case of the particle inside the electromagnetic field, we are dealing
with the Galilei group extended by $U(1)(\vec{x},t)$, that is, 
$\varphi=\varphi(\vec{x},t)$. This algebra is infinite-dimensional but,
for analytical functions $f$, we can resort to the following economical
short cut. We start by adding to Galilei algebra only those generators 
$f\otimes \tilde{X}_\varphi$ for which $f$ are linear functions, i.e. 
$t\otimes \tilde{X}_\varphi$ and $x^i\otimes \tilde{X}_\varphi$, to be referred
as $\tilde{X}_{A^0}$ and $\tilde{X}_{A^i}$, respectively. Let us call 
$\tilde{G}_E$ the group 
associated with this finite-dimensional algebra.
The commutation relations of $\tilde{\cal G}_E$ are (omitting zero
commutators as well as rotations, which operate in the standard way):
\be
\begin{array}{rclrcl}
{}[\tilde{X}_{v^i},\,{\tilde{X}_t}]&=&\tilde{X}_{x^i} & 
[\tilde{X}_{x^i}\,,\tilde{X}_{v^j}]&=& -m\delta_{ij}\tilde{X}_\varphi\\
{}[\tilde{X}_t,\,\tilde{X}_{A^0}]&=&-q\tilde{X}_\varphi & [\tilde{X}_{x^i},\,
\tilde{X}_{A^j}]&=& q
\delta_{ij}\tilde{X}_\varphi \\
{}[\tilde{X}_{v^i},\,\tilde{X}_{A^i}]&=&\delta_{ij}\tilde{X}_{A^0} & {}\nn
\end{array}\label{Electro}
\ee
\ni where we have performed a {\it new central extension}, which is allowed
by the Jacobi identity and parametrized by $q$. This parameter 
will eventually be identified (see bellow) with the electric charge of 
the particle. 

Fortunately, $\tilde{G}_E$ encodes all the dynamical information in the
local group. The only effect of including all functions $f$ is that of 
making local the group parameters $A^0$ and $A^i$ (corresponding
to $\tilde{X}_{A^0}$ and $\tilde{X}_{A^i}$, repectively), something we 
shall recover here exactly by the imposition of an appropiate constraint.
After the exponentiation of the group, we compute the quantization one-form
which turns out to be:
\bea
\Theta=m\vec{v}\cdot d\vec{x}+q\; \vec{A}
\cdot d\vec{x}-(\frac{1}{2}m\vec{v}^2+q 
A_0)dt+d\varphi \ \ ,
\eea
which is again the Poincar\'e-Cartan of a particle inside a electromagnetic 
field plus the central term $d\varphi$. In order to derive the equations of 
motion of the particle we have to impose
the above-mentioned constraint, making the functions $A^\mu$ ($A^i$ and $A^0$)
to depend on the position of the particle: $A=A(x_
{particle})$. The vector field $X$ in the kernel of $\Theta$, that is,
satisfying  $i_Xd\Theta=0$ is:
\bea
X=\frac{\partial}{\partial t}+\vec{v}\cdot\frac{\partial}{\partial \vec{x}}
-\frac{q} {m}[(\frac{\partial A_i}
{\partial x^j}-\frac{\partial A_j}{\partial x^i})v^j+\frac{\partial A_0}
{\partial x^i}+\frac{\partial A_i}{\partial t}]\frac{\partial}{\partial v_i}
\eea
and its trajectories are governed by the following equations:
\be
\begin{array}{l}\frac{d\vec{x}}{dt}=\vec{v} \\
m\frac{d\vec{v}}{dt}=q
[\vec{v}\wedge(\vec{\nabla}\wedge\vec{A})-\vec{\nabla}A_0-\frac{\partial 
\vec{A}}{\partial t}] \end{array}\label{eleqmot}
\ee
we results in standard expression for the Lorentz force,
when we define $\vec{\nabla}\wedge\vec{A}\equiv\vec{B}$ and 
$-\vec{\nabla}A_0-\frac{\partial \vec{A}}{\partial t}\equiv
\vec{E}$.
We have studied this electromagnetic example in the Galilean scheme for
pedagogical reasons, but we must stress that everything can be
reproduced in the relativistic case, starting from the Poincar\'e group
and constructing the corresponding $\tilde{\cal P}_E$, finally leading to
the same final expression.

\subsection{Electromagnetism and gravity mixing}
We address now a more involved system using 
for its analysis the 
most algorithmic of the two techniques we have presented till now: GAQ.

We start directly with the centrally extended Poincar\'e group and try to
make local the space-time translation subgroup, 
instead of the central $U(1)$ one. These 
local translations can be seen as local diffeomorphisms, thus suggesting
the emergence of gravity notions \cite{Kibble}.
The Lie algebra can be written as:
\bea
\left[\tilde{X}_{t},\tilde{X}_{v^i}\right]=\tilde{X}_{x}\;\;\;,\;\;\;
\left[\tilde{X}_{x^i},\tilde{X}_{v^j}\right]=-\delta_{ij}(\tilde{X}_{t}+
m \tilde{X}_\varphi) \ \ . 
\eea
An interesting phenomenon shows up when we make the space-time 
translations local. In fact, when computing the commutators following 
the general rule we gave above, we find,
\bea
[\tilde{X}_{v^i},f\otimes \tilde{X}_{x^j}]=(X_{v^i}f)\otimes \tilde{X}_{x^j}+
\delta_{ij}(f\otimes \tilde{X}_{t}+f\otimes \tilde{X}_\varphi)\;\;, 
\eea
This means that making local the translation generators in the extended 
Poincar\'e group implies the appearance of a local $U(1)$ symmetry. This fact
is linked to the loss of invariant character of the translation subgroup
in the extended Poincar\'e group. As we saw in the 
previous subsection, making local the central term leads to a coupling 
between the 
particle and an electromagnetic force. Therefore, we find that introducing
the gravitational field offers the quite interesting possibility of an 
automatic coupling of the particle and an electromagnetic field and 
suggests the possibility of a mixing between both interactions.

In order to derive the corresponding dynamics we undertake exactly the same 
path we follow in the pure electromagnetism case. 
(Symmetrized) Generators of local space-time translations associated with 
the linear functions, $x_\mu\otimes{P}_\nu+x_\nu\otimes P_\mu$, 
will be called
$X_{h^{\mu\nu}}$, and close the finite-dimensional
algebra $\tilde{\cal P}_{EG}$ ($\supset\tilde{\cal P}_E$) analogous to 
$\tilde{\cal G}_E$.
The analogue of the vector $A^\mu$ is now a symmetric tensor $h^{\mu\nu}$, 
eventually interpreted as a metric. 
However, the central-charge structure
of this finite-dimensional electro-gravitational subgroup, $\tilde{P}_{EG}$, is
richer than that of $\tilde{P}_E$ or $\tilde{G}_E$, entailing a more
involved GAQ analysis which  must be made, for the time being at 
least, order by order. For the sake of clarity, we shall dwell here on a
simplified case corresponding to a {\it non-relativistic} limit, which still 
preserves the main features of the discussion.
This limit is achieved  by means of an In\"on\"u-Wigner 
contraction of $\tilde{\cal P}_{EG}$  with respect to the subalgebra  
spanned by $\tilde{X}_t, \tilde{X}_{A^i}$ and the rotations.
The contracted algebra shows (omitting rotations):
\begin{equation}
\begin{array}{rclrcl}
\left[\t,\tilde{X}_{v^i}\right]&=&P_i &\left[\t,\tilde{X}_{h^{00}}\right]&=& 
-2m \tilde{X}_\varphi \\
\left[\t,\tilde{X}_{h^{0i}}\right]&=&\tilde{X}_{x^i} &\left[\t,
\tilde{X}_{A^0}\right]&=&q\tilde{X}_\varphi \\
\left[\tilde{X}_{x^i},\tilde{X}_{v^j}\right]&=&-(m+\kappa q)\delta_{ij}
\tilde{X}_\varphi &\left[\tilde{X}_{x^i},\tilde{X}_{A^j}\right]
&= &-q\delta_{ij}\tilde{X}_\varphi\\
\left[\tilde{X}_{x^i}, \tilde{X}_{h^{0j}}\right]&=&m\delta_{ij}
\tilde{X}_\varphi &
\left[\tilde{X}_{v^i},\tilde{X}_{h^{0j}}\right]&=&
-\delta_{ij}\tilde{X}_{h^{00}}+\kappa\delta_{ij}\tilde{X}_{A^0}  \\
\left[\tilde{X}_{v^i},\tilde{X}_{A^j}\right]&=&-\delta_{ij}\tilde{X}_{A^0} & 
 \left[\tilde{X}_{h^{0i}},\tilde{X}_{A^j}\right]&=&
-\delta_{ij}\tilde{X}_{A^0}  
\end{array}
\end{equation}
The most significant characteristic of this algebra is the appearance of a 
new constant $\kappa$ associated with the already commented mixing of 
electromagnetic and gravity forces (this fact is apparent in the
non-contracted algebra where commutators of the type 
$[\tilde{X}_{h^{\mu\nu}},\tilde{X}_{h^{\alpha\beta}}]\sim\tilde{X}_{A^\rho}$ 
are present). The appearance of this $\kappa$ in the commutator 
between boosts and translations in fact modifies the inertial mass by a term
$\kappa q$, something most relevant from the physical point of view.

The next step requires the exponentiation of the algebra and the 
subsequent construction of the quantization
one-form from which the equations of motion can be derived.
The exponentiation process is still complicate and we employ a consistent 
order by order procedure for the which can be found 
in \cite{exponen}.
The following equations of motion correspond to the first non-trivial
terms approximating the complete equations:
\bea
\frac{d\vec{x}}{dt}&=&\vec{v}
\nn \\
(m+\kappa q)\frac{d\vec{v}}{dt}&=&q\left[\vec{v}\wedge(\vec{\nabla}\wedge
\vec{A})-\vec{\nabla}A^0-\frac{\partial\vec{A}}{\partial t}\right] \nn \\
&-& m\left[\vec{v}\wedge(\vec{\nabla}\wedge
    \vec{h})-\vec{\nabla}h^{00}-\frac{\partial\vec{h}}{\partial t}\right]+
\frac{m}{4}
    \nabla(\vec{h}\cdot\vec{h}) \label{egeq}\\     
&-&\frac{\kappa q}{2}\left[\vec{v}\wedge(\vec{\nabla}\wedge\vec{h})-\frac{1}{4}
\nabla(\vec{h}\cdot\vec{h})-\frac{\partial\vec{h}}{\partial t}\right]
\nn
\eea
On the kinematical part (l.h.s.), we explicitly notice what was 
already foreseen at the Lie-algebra level, i.e. the kinematical mass is 
corrected by a term proportional to $\kappa$ and the charge of the particle.

\ni On the dynamical side, the first line  is again the expression of the 
Lorentz force, while the expression in the second line, known as 
{\it  gravito-electromagnetism} \cite{Wald}, corresponds to the 
{\it geodesical motion}
in its first non-trivial perturbative expression (linearized gravity), which
is the one obtained when working in the group law up to the third order in 
group variables as we have done.
The last line is proportional to the mixing parameter $\kappa$ and shows
the appearance of a new force of {\it electromagnetic} behaviour but of 
{\it gravitational} origin, consequence of this new possibility opened by the
analysis of the underlying symmetry cohomlogy.

\section{Conclusions}

We have seen how physical constants characterizing the particle and its 
couplings arise from the parameters associated with the cohomology of the 
symmetry underlying the physical system. This was already well-known for the
mass $m$ and it is explicitly shown here for the electric charge $q$.

We have recovered in an algebraic setting two standard interactions, 
electromagnetism and gravity, by making local some invariant subgroups in the
{\it kinematical} symmetry of the particle. This in fact constitutes 
a revision of the gauge principle in this framework.
When exploring the full possibilities that the Lie algebra permits,
a new force parametrized by a constant $\kappa$ has also be found  
associated with a mixing of the standard previous interactions. 
Nature could choose $\kappa$ to be zero, but the presence if this new
term is a possibility that the Lie algebra definitely offers.
A crucial observation of this mixing process is the relevance of
making local the symmetry (translations) {\it after} a central extension 
of the group has already been performed. This endows this new interaction 
with a {\it quantum flavour} since such a central extension 
is intrinsicly tied to the quantum symmetry of the
corresponding system \cite{Michel}.

A value of $\kappa$ different from zero has, in principle, two direct testable 
consequences. Firstly, it entails a $2\kappa q$ mass difference between 
charged particles and anti-particles. An algebraic treatment like the
present one, is not primarily related to additional physical phenomena such
as radiative corrections, but it offers in turn a conceptual algebraic 
framework in the case these effects actually occur.
In that case, the current experimental clearance  in 
the values of mass differences in pairs like electron-positron represent an 
upper bound  for the constant $\kappa$, implying a very small value (around
$10^{-8}m_e$). Nevertheless, this tiny value would have strong and fundamental 
implications, specially the violation of CPT symmetry. 

\ni Secondly,  since $\kappa$ modifies the inertial mass (l.h.s. in 
(\ref{egeq})) but leaves untouched the gravitational mass (in the (r.h.s)) 
it represents an explicit violation of the {\it weak equivalence principle},
a result with far-reaching conceptual consequences.

\end{document}